\documentclass[]{interact}

\usepackage{epstopdf}
\usepackage[caption=false]{subfig}
\usepackage{amsmath,amsfonts}
\usepackage{algorithmic}
\usepackage{array}
\usepackage{caption}
\usepackage{subcaption}
\usepackage{float}
\usepackage{placeins}
\usepackage{bm}
\usepackage{xcolor}
\usepackage{adjustbox}
\usepackage{amssymb}

\theoremstyle{plain}

\theoremstyle{definition}

\theoremstyle{remark}

\begin{document}

\articletype{ARTICLE TEMPLATE}

\title{Electric Vehicle Charging Stations Placement Optimization in Vietnam Using Mixed-Integer Nonlinear Programming Model}

\author{
\name{
    Vu Truc Quynh\textsuperscript{a}, 
    Ha Hien Minh\textsuperscript{a}
    and Vu Tuan Hai\textsuperscript{b}\thanks{Email: vu.tuan\_hai.vr7@naist.ac.jp}}
\affil{
    \textsuperscript{a}Foreign Trade University, Ho Chi Minh City Campus, Vietnam; \\
    \textsuperscript{b}Nara Institute of Science and Technology, 8916–5 Takayama-cho, Ikoma, Nara 630-0192, Japan}
}

\maketitle

\begin{abstract}
Vietnam is viewed as one of the promising markets for electric vehicles (EVs), especially automobiles, when it is predicted to reach 1 million in 2028 and 3.5 million in 2040. However, the lack of charging station infrastructure has hindered the growth rate of EVs in this country. This study aims to propose an optimization model using Mixed-Integer Nonlinear Programming to implement an optimal location strategy for EVs charging stations in Ho Chi Minh City. The problem is solved by Gurobi using the Brand-and-Cut method. There are two perspectives, including Charging Station Operators and EV users. In addition, 7 kinds of costs are considered. From 1509 Point of Interest and 199 residential areas, 134 POIs were chosen with 923 charging stations to fully satisfy the customer demand. Furthermore, the effectiveness of the proposed model is proved by a minor MIP Gap and running in a short time with full feasibility.
\end{abstract}

\begin{keywords}
Electric vehicles, Charging station, Mixed-Integer Nonlinear Programming, Optimization.
\end{keywords}

\section{Introduction}

In the new global economy, environmental issues have become an important worldwide problem. Transport sector accounts for magnification of this, with millions of gasoline-powered vehicles on the roads emitting carbon continuously, carbon emissions persist at a high rate. Besides, gasoline-powered cars also create cacophonous pollution from the internal combustion engine \cite{yang2020evaluation}. In order to alleviate these vulnerabilities and promote a more sustainable economy, one solution is to switch from gasoline-based vehicles to using green technology vehicles. For Vietnam, these advantages align closely with the country’s sustainability objectives, helping move it closer to its Net Zero goal by 2050. 

The Vietnamese market is regarded as one of the fastest-growing, with the domestic sector projected to achieve the significant milestone of 1 million EVs, then increasing to 3.5 million by 2040 \cite{vama}. Despite this promising growth, the current share of EVs remains relatively low compared to global levels. The lack of charging infrastructure poses a significant challenge for Vietnam’s EV industry, as the limited availability of charging stations (stations) dampens consumer confidence and hinders the practical adoption of EVs. Although 70\% of respondents expressed interest in purchasing an EV, a significant portion remains hesitant to proceed. The primary factor influencing this indecision, affecting 18\% of respondents, is the limited availability of accessible stations nationwide \cite{kpmg2024}. This shortage highlights a critical need for infrastructure expansion to support the growing interest in EVs.

Developing a strategic model for planning EV stations has become crucial, as a well-organized distribution of stations can optimize available resources, ultimately reducing investment costs for both investors and the Government. Furthermore, such a model would enable a comprehensive analysis of long-term costs and benefits. This approach accelerates the growth of EVs and contributes to reducing dependence on fossil fuels and enhancing energy efficiency.  Conversely, the absence of a clear planning framework could lead to an uneven distribution of stations, resulting in resource wastage and reduced system efficiency. Thus, the objective of this paper is to effectively promote the placement of EV stations in the rapidly growing EV market, the application of an optimization model in planning and establishing these stations is indispensable.

The contributions of this paper can be summarized in two key aspects. 

\begin{itemize}
    \item This study simultaneously explores the perspectives of both station operators and users, proposing solutions that balance the interests of providers and consumers.
    \item It evaluates the economic aspects of stations within a new survey scope (HCM City).
\end{itemize}

As a result, the analysis identifies Thu Duc City as having the highest POIs (19.02\%) and population (13.06\%), highlighting its strong potential for EV charging infrastructure. High demand is also observed in District 12, District 7, and Binh Thanh District, while suburban areas like Can Gio District and Nha Be District require infrastructure expansion despite lower demand. Newly installed stations are concentrated in central urban and underserved suburban areas, with Level-2 stations generally outnumbering Level-3 stations, except in Binh Chanh District and Thu Duc City. Charging costs dominate total expenses (35.47\%), followed by operational costs (28.83\%) and travel costs (27.99\%).


\section{Literature Review}

\begin{figure*}
    \centering
    \caption{Compiles of main research directions about EV placement}
    \includegraphics[width=0.90\linewidth]{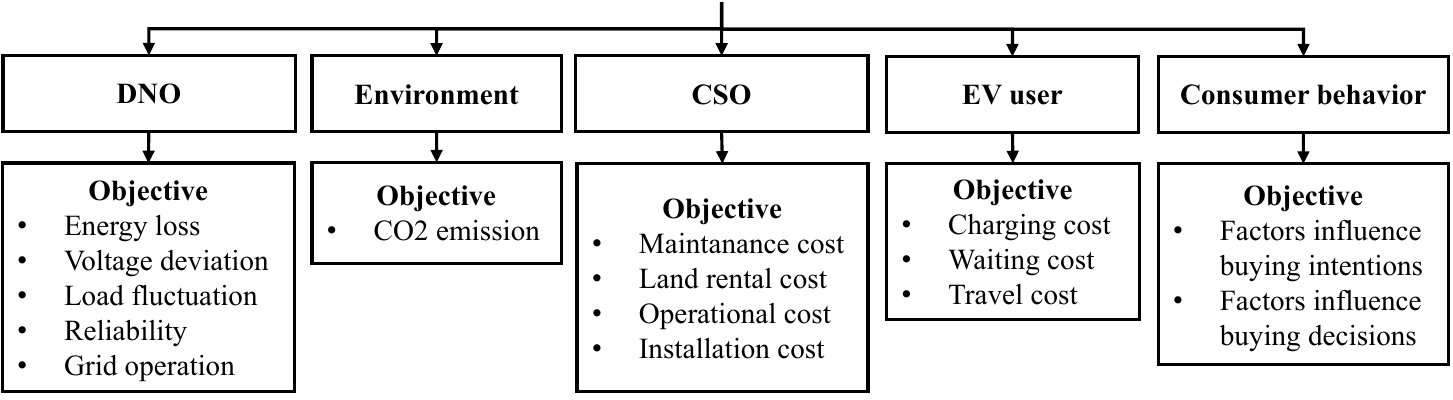}
    \label{fig:direction}
\end{figure*}

\begin{table*}[]
\caption{Summary of previous research}
\resizebox{0.99\textwidth}{!}{%
\begin{tabular}{|p{1.0cm}|p{4cm}|p{3cm}|p{3.5cm}|p{5.2cm}|}
\hline
\textbf{Ref} & \textbf{Objective function} & \textbf{Model} & \textbf{Metrics} & \textbf{Dataset} \\ \hline 
 \multicolumn{5}{|c|}{\textbf{DNO directions}}\\ \hline 
 \cite{nugra22}& Minimizing power loss, network capacity and voltage regulation& HGAMSSA & Convergence process & IEEE 33 bus distribution network\\ \hline
\cite{Khalil22}&                    Balancing electrical resistance&       LP&         Optimal Gap, Time Consumption&         IEEE 69 Bus System\\ \hline
 \cite{Zhang22}&  Enhancing the overall stability of the power grid& GSA-PSO hybrid optimization& Average calculation time, Optimal value, Average value&None\\\hline
 \cite{tan23}& Renewable energy sources beyond the grid& PVSOL, AutoCAD& Cost, Payback period&None\\\hline
 \cite{mena21}&  Charger design for greater quality and performance& Input-Series Output Parallel (ISOP) DC-DC converter&IVS, ICS, and OVS&None\\\hline
 \cite{quang23}& Reducing battery losses in EVs& Dynamic Programming& Energy loss&None\\ \hline 
 \multicolumn{5}{|c|}{\textbf{Environmental direction}}\\\hline
 \cite{lu17}& Reducing greenhouse gas emissions associated with electricity production and transmission used for charging EVs& Nonlinear programming using PSO algorithm& Average calculation time, Optimal value, Average value&None\\ \hline 
 \multicolumn{5}{|c|}{\textbf{CSOs direction}}\\\hline
 \cite{sun23}& Optimizing maintenance costs& SA-IPSO & Solution time & IEEE33\\\hline
 \cite{raga16}& Optimizing land rental costs& Linear programming (MDR model, multi-server queue)& Demand satisfied, Minimum required budget, Feasibility gap&UK’s National Charge Point
Registry data, OpenStreetMap\\\hline
 \cite{moha24}&  Optimizing operational costs& STM (PSO algorithm)& PDFs, Economic/environmental feasibility&Hainan Statistical Yearbook 2014\\\hline
 \cite{cui19}& Optimizing installation costs& MILP& Average minimum total cost, Computational time&IEEE 123-bus distribution system\\ \hline 
 \multicolumn{5}{|c|}{\textbf{EV Users direction}}\\\hline
  \cite{zafa21}& Maximum Coverage Location Problem (MCLP)& LP (PuLP with GLPK package)&Wake County Government GIS, Google Maps & None\\\hline
 \cite{bano20}&  Assist the grid during emergencies & FCR-N (local search algorithm)& Economic cost&None\\\hline
  \cite{nimal21}&  Manage charging and discharging during off-peak and peak hours& N-EVC(D)& Economic cost, Voltage fluctuation&IEEE 13 node test feeder (Australia)\\\hline
 \cite{zhou22}& Optimizing charging fees & Linear programming (Genetic Algorithm) & Total social cost & SRI, SEAI, CSO (Ireland), Wikipedia\\\hline
 \cite{chen20}& Optimizing waiting time costs& MINLP model (LINGO, GA and NGA)&  Total hourly cost & Real network in Xi’an China \\\hline
  \cite{Kong19}& Optimizing travel costs& Dijkstra
routing method& Economic cost, Charging waiting time, Average travelling velocity & The Special Plan for Electric Vehicle Charging Infrastructure in Beijing (2016 - 2020), Beijing Traffic Development Annual Report\\\hline
 \multicolumn{5}{|c|}{\textbf{Consumer behavior direction}}\\\hline 
 \cite{dk19}&  Perceived risks—namely financial, functional, and social risks—impact purchasing decisions& TAM& None&None\\\hline
\end{tabular}%
}
\label{tab:related-work}
\end{table*}

A large number of research has proliferated and can be summarized into five perspectives: DNOs, environment approach, CSOs, EV users, and consumer behavior. These research directions are summarized in the Figure~\ref{fig:direction} and Table ~\ref{tab:related-work}. Furthermore, five approaches are elaborated below.

Research on the entity responsible for managing and operating the electrical power distribution network or Distribution Network Operators (DNOs) focuses on optimizing the operation and management of electricity networks. Studies in this field aim to improve the operational efficiency of stations by addressing engineering challenges, such as minimizing voltage fluctuations \cite{nugra22}, balancing electrical resistance \cite{Khalil22}, and enhancing the overall stability of the power grid \cite{Zhang22}. Additionally, research is also directed at meeting high levels of consumer demand through renewable energy sources beyond the grid \cite{tan23}, improving charger design for greater quality and performance \cite{mena21}, and reducing battery losses in EVs \cite{quang23}. These efforts are essential to support the growing demand for sustainable energy solutions in EV infrastructure.

Studies on the environmental impact of vehicles highlight key differences between conventional vehicles and EVs. Internal combustion engine (ICE) vehicles emit pollutants directly through exhaust pipes, as well as through evaporation from fuel systems and during refueling. In contrast, fully EVs produce no direct emissions. Due to challenges in accurately measuring emissions from EVs, in-depth research on the environmental impact remains limited \cite{ray23}. Consequently, rather than focusing on direct emissions from EVs, most studies in this area emphasize reducing greenhouse gas emissions associated with electricity production and transmission used for charging EVs \cite{lu17}.

Research focusing on the perspective of Charging Station Operators (CSOs) highlights the role in managing and ensuring the daily operation of stations. These operators are responsible for investing in, installing charging equipment, and monitoring the revenue generated from these stations. Additionally, they purchase electricity from local and regional energy distribution networks then deliver it to end users by coordinating with DNOs during the installation process. CSOs can range from individuals and households to businesses. According to \cite{aayog21}, most CSOs in the Indian market are individuals. Unplanned installations, however, can have adverse effects on the overall system, particularly in terms of costs\cite{ray23}. Consequently, the primary goal of CSOs is to identify locations that maximize profitability while minimizing costs, all within the constraints of policies, economics, and infrastructure. Research in this area focuses on optimizing costs associated with station operations, including maintenance costs \cite{sun23}, land rental costs \cite{ahmad22}, operational costs \cite{moha24}, and installation costs \cite{cui19}.

Research related to EV users, the owners who are responsible for recharging their vehicles at these stations. Since most stations have listed prices, the primary concern for EV owners is the distance they must travel to reach a station. Studies on problems like the Maximum Coverage Location Problem (MCLP) \cite{zafa21}. Additionally, in countries that support and implement Vehicle-To-Grid (V2G) technology - which enables energy to flow back from EVs to the grid - research aims to optimize the timing and battery percentage that station users adjust to assist the grid during emergencies or manage charging and discharging during off-peak and peak hours \cite{bano20}, \cite{nimal21}. The overarching objective of these studies is to strategically place stations in locations that minimize travel costs while effectively meeting the charging demands of surrounding areas. Furthermore, optimizing the costs associated with stations - including charging costs \cite{zhou22}, waiting time costs \cite{chen20}, and travel costs \cite{Kong19} - is a significant focus in this research field.

Studies on consumer behavior focus on analyzing the factors that directly motivate consumers to consider purchasing EVs or to transition from traditional gasoline cars to this novel mode of transportation.\cite{dk19} expanded the Technology Acceptance Model (TAM) to explore how perceived risks - namely financial, functional, and social risks - impact purchasing decisions. Meanwhile, other studies have highlighted various factors such as perceived benefits, ease of use, facilitating conditions, cost perception, usage habits, and environmental awareness, perceived usefulness, ease of use, social influence, social barriers, charging infrastructure, and environmental awareness in shaping consumer decisions. In summary, those studies regarding consumer behavior provide insights into the motivation and factors influencing customer decisions when shifting towards EVs.

In Vietnam, since the electric vehicle industry has only recently emerged, related research has primarily focused on technical issues such as electrical load, reliability, batteries, and consumer behavior in purchasing decisions. Conversely, economic aspects related to EVs remain unexplored, presenting a new area for further investigation. Therefore, focusing on DNOs, CSOs, and EV users offers promising research directions. Regarding the placement of charging stations to optimize economic benefits for grid operators (DNOs), according to \cite{qh15}: “The State (Vietnam Government) holds a monopoly over national electric system regulation.” This means that in Vietnam, the Government retains exclusive control over the electricity system, particularly transmission, making studies focused on grid operation challenging to apply within the Vietnamese context. At the same time, existing studies on EV charging station placement offer diverse perspectives; however, they are often based on simulated data \cite{nugra22} or tailored to specific regions such as the UK \cite{raga16} and China \cite{moha24}. These models may not be directly applicable to Ho Chi Minh City, given its unique geographical characteristics and public policy framework. Therefore, this research adapts existing methodologies to better align with the local context, ensuring a more practical and region-specific approach. Building on the Vietnamese context and insights from international studies, this research aims to incorporate previous findings and address existing research gaps.

\section{Backgrounds}
\label{sec:background}

\begin{table}[h!]
\caption{Notation and definitions used in the MINLP model}
\label{tab:notation_definitions}
\centering
\renewcommand{\arraystretch}{1.1} 
\begin{adjustbox}{center, max width=\textwidth}
\begin{tabular}{|p{0.12\textwidth}|p{0.85\textwidth}|} 
\hline
\textbf{Notation} & \textbf{Definition} \\ \hline

$x_{2,i}$& $\#\text{station}_{\text{Level}-2}$ at location $i $\\ \hline
$x_{3,i}$& $\#\text{station}_{\text{Level}-3}$ at location $i $\\ \hline
$n$ & $\#$POI\\ \hline
$i_{2,i}$& The installation cost of $\text{stations}_{\text{Level}-2}$ at location $i $\\ \hline
$i_{3,i}$& The installation cost of $\text{stations}_{\text{Level}-3}$ at location $i $\\ \hline
$l_i$ & The land rental cost of location $i $\\ \hline
$m_{2,i}$& The maintenance cost of $\text{stations}_{\text{Level}-2}$ at location $i $\\ \hline
$m_{3,i}$& The maintenance cost of $\text{stations}_{\text{Level}-3}$ at location $i $\\  \hline
$e_{2,i}$& The energy consumption of $\text{stations}_{\text{Level}-2}$ at location $i $\\  \hline
$e_{3,i}$& The energy consumption of $\text{stations}_{\text{Level}-3}$ at location $i $\\ \hline
 $\overline{e}$&The average amount of electricity needed to power a vehicle for one day \\ \hline
 $\overline{P_o}$&The average electricity cost that a CSO has to pay per 1 kWh\\ \hline
 $\overline{P_u}$&The average charging cost that an EV user has to pay per 1 kWh\\ \hline
$\overline{P_{km}}$& The average charging cost that an EV user has to pay per 1 km\\ \hline
$\overline{w}$& The average wage per capita in HCM City\\ \hline
$r$& Traffic rate of a station\\ \hline
$d_{i,j}$& Distance from the RP $i $ to station $j$\\ \hline
$B_{HCM}$& The budget for building stations in HCM City\\ \hline
$d_{max}$& The max distance that an electric automobile can travel\\ \hline
$c_{i}$& The \#vehicles in residential area $i$.\\ \hline

\end{tabular}
\end{adjustbox}
\end{table}

Ho Chi Minh City has been selected as the ideal location for developing an optimized electric vehicle (EV) charging station model due to its large population and the increasing demand for EVs. Despite covering only 0.63\% of the country’s total area, with a land size of 2,095 km², the city had a population of over 9 million as of December 2022, accounting for nearly 10\% of Vietnam’s total population \cite{ctkhcm23}. This makes it not only the most populous city in Vietnam but also the country’s economic hub, characterized by a dynamic urban lifestyle and a high volume of traffic. Given these factors, Ho Chi Minh City presents a compelling case for implementing an optimized EV charging station model to support sustainable urban mobility.

The problem of optimizing EV stations involves numerous variables, classifying it among highly complex (NP-hard) problems. To date, various methods have been employed to address these optimization challenges, including mathematical optimization and model-based programming techniques. The notation, as presented in Table~\ref{tab:notation_definitions}, is presented throughout the MINLP model.

\subsection{Mixed-Integer Nonlinear Programming Model}

In this research, we choose the Mixed-Integer Nonlinear Programming model (MINLP) which is widely used in recent studies \cite{batta19,afs19,kim20}, as it integrates linear, nonlinear, and mixed-integer programming. This technique can handle both continuous and discrete variables along with nonlinear functions, providing reliable solutions for multi-objective problems. The MINLP model is particularly suitable for handling complex problems that involve nonlinearity and integer variables, often requiring extensive search trees for solutions. Specifically, the target is that minimizing objective function $f(x)$ with constraint $c(x)$:

\begin{align}
    \text{min}\;f(x)\;\text{s.t}\;c(x) \leq 0, \; x \in X, \; x_i \in \mathbb{Z} \; \forall i \in I,
\end{align}

where $x\in\mathbb{R}^n$ and constraint function $ c: \mathbb{R}^n \to \mathbb{R}$. Both $f$ and $c$ must be twice differentiable and well-defined. Additionally, $X$ is a convex polyhedral space, while $ x_i \in \mathbb{Z} $ for $ I \subseteq \{1, \ldots, n\} $, with $I$ as a subset (or equal) of integer indices starting from 1 onward. For extended models, the objective function can involve maximization or constraints defined by bounds $\{l,u\}$, such as $l \leq c(x) \leq u$.

\subsection{Branch and Cut Algorithms}

MINLP can be divided into two cases based on the objective function: convex or concave. For the convex case, two solution methods are commonly used: the single-tree approach, which includes ``Branch and Bound'' and ``Branch and Cut'' methods; and the multi-tree approach including outer approximation and Benders decomposition. Among these, the Branch and Bound method, invented in 1960 \cite{land60}, is particularly noteworthy. This method transforms the original MINLP model into a Mixed-Integer Nonlinear Programming (MILP) model, making the problem more tractable \cite{zhang20}. 

\begin{figure}
    \centering
    \includegraphics[width=0.7\linewidth]{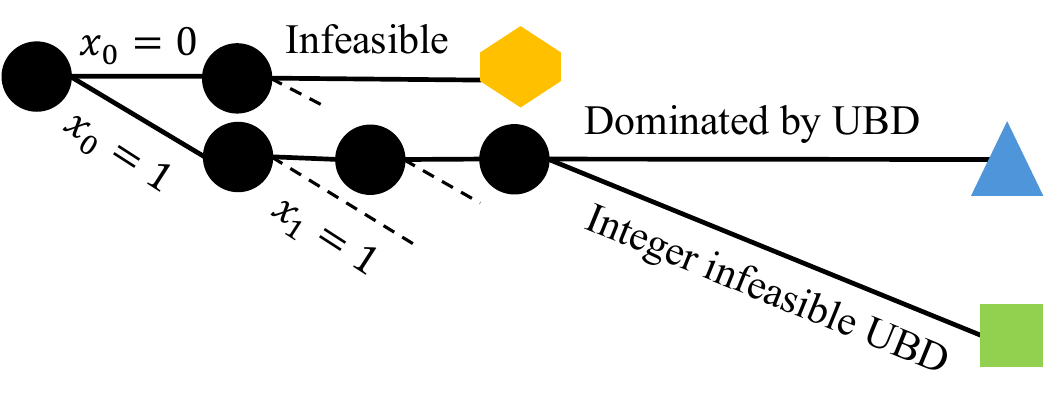}
    \caption{The branch is cut as a dotted line because of an infeasible solution. The normal line - branch continues to develop to achieve the best solution until dominated by the upper bound (UBD).}
    \label{fig:bba}
\end{figure}

Since the goal is to find the optimal solution, $f(x)$ is optimized to reach the best possible value. To achieve this, the algorithm iteratively subdivides the main problem into smaller ``branches'' (sub-problems) and uses constraints (``bounds'') to either retain or discard solutions, continuously updating the upper and lower bounds to narrow down feasible solutions. The improved version, known as Branch and Cut \cite{gomo58}, is widely used to solve complex problems by integrating Cutting Planes, which are additional constraints dynamically generated during the solving process. These cuts help reduce the feasible region by eliminating fractional or non-integer solutions, thereby narrowing the search space and reducing the size of the branch-and-bound tree. This makes Branch and Cut significantly more efficient for solving complex problems, especially in large-scale integer programming, compared to the traditional Branch and Bound approach. Fig.~\ref{fig:bba} presents the abstract idea of the Branch and Cut method by illustrating the branching process in Branch and Bound method step-by-step and eliminating infeasible branches or branches that do not lead to optimal solutions. Each branching step reduces the search space, thereby enhancing efficiency and reducing solution time. 

The problem can be formulated as follows: given a point $ x^{(l,u)} $ where $ x^{(l,u)} \notin \{0,1\} $ and a feasible set $ F^{(l,u)} $ associated with that point, the goal is to find a vector $ \left( \pi_0, \pi^T \right)^T $ such that the inequality $ \pi^T x \leq \pi_0 $ holds for all $ x \in F^{(l,u)} $. The algorithm simultaneously excludes values of $ x^{(l,u)} $ that satisfy $ \pi^T x > \pi_0 $. The large MINLP problem is then broken down into multiple smaller Nonlinear Programming (NLP) sub-problems (nodes) using relaxation techniques.

\section{Methodology}
\label{sec:method}

\subsection{Variables}
\label{sec:var}
According to the SAE J1772 standard, which is widely adopted by EV manufacturers and stations globally, including in Vietnam, stations are categorized into three levels based on charging power, with \textbf{Level-2} and \textbf{Level-3} chargers being the focus of this study. Level-2 chargers, powered by single-phase or three-phase AC, offer charging capacities between 3.7 kW and 22 kW, making them ideal for home, workplace, and public locations such as parking lots and office buildings. On the other hand, Level-3 chargers, also known as DC fast chargers, bypass the onboard AC/DC converter and directly charge the vehicle's battery with DC power at capacities exceeding 44 kW. They are capable of replenishing up to 80\% of an EV's battery in just 20 - 30 minutes, making them felicitous for highway rest stops or other short-duration public stops \cite{hana21}. Level-1 chargers are compact emergency units, but their limited power capacity leads to long charging times, making them suitable only for home use or wall mounting. Furthermore, they typically lack user data communication capabilities \cite{kim23}. Due to these limitations and their unsuitability for public use, this study will not consider Level-1 stations. In short, $x$ in our research is comprised of of $x_{2,i}$ and $x_{3,i}$, namely $x$ = $x_{2,i}$ + $x_{3,i}$.

\subsection{Objective function}
From the investors' perspective, according to utility theory \cite{neumann1944}, their primary goal is to minimize budget expenditures as much as possible during the implementation of stations. Consequently, minimizing total costs as \eqref{eq:basic}:

\begin{equation}
\begin{split}
  \mathcal{C}(x)=\sum_{i=1}^{n} (I_i + L_i + M_i 
+ O_i + C_i + W_i + T_i) 
\end{split}    
\label{eq:basic}
\end{equation}

across $n$ locations when installing and operating a station from CSOs and EV users perspectives, taking into account 7 cost components associated with the deployment and operation of stations. Table~\ref{tab:factor} summarizes the definitions and meanings of the cost components denoted in~\eqref{eq:basic}.

\begin{table*}[]
\caption{Summary of 7 different components}
\resizebox{0.99\textwidth}{!}{%
\fontsize{10}{12}\selectfont
\begin{tabular}{|p{1.2cm}|p{1.85cm}|p{10.7cm}|p{3.2cm}|}
\hline
\textbf{Symbol}& \textbf{Stand for}& \textbf{Meaning}& \textbf{Formula}\\ \hline
$I_i$ & Installation cost & Covers the initial setup required at each station site $i$, its components are  charging stations, charging cables, and related hardware. For fixed costs, such as installation costs, the author uses a straight-line depreciation calculation method for budgeting purposes.& $\sum_{k\in\{2,3\}}x_{k,i} I_k$ \\ \hline
$L_i$ & Land rental cost & The expense of securing space for the station daily. & $\frac{l_i}{14600} (x_{2,i} + x_{3,i})$ \\ \hline
$M_i$ & Maintenance cost & The expenses incurred by station operators during the operation of electric charging units to ensure their continuous functionality. According to \cite{batta19}, maintenance costs encompass the upkeep of transformers, charging equipment, and other devices installed at charging sites. In this study, the author collectively refers to all costs associated with the maintenance of equipment at stations as maintenance costs. & $\sum_{k\in\{2,3\}}x_{k,i} M_k$ \\ \hline
$O_i$ & Operation cost & The electricity expense that the CSOs must pay to HCM City Power Corporation. As previously mentioned, the Government holds exclusive control over electricity transmission, while Vietnam Electricity monopolizes distribution and has the authority to set electricity prices in the country. According to \cite{deb21}, the operational cost can be calculated as a function of the electricity used, multiplied by the total $\#$stations in operation and the electricity rate. & $(\sum_{k\in\{2,3\}}x_{k,i} e_k)\overline{P_o}$ \\ \hline
$C_i$ & Charging cost &The average electricity fee that the EV users must pay to the CSO. Charging costs are calculated as the product of electricity price, power consumption per hour, and the number of hours of consumption \cite{zhou22}. & $(\sum_{k\in\{2,3\}}x_{k,i} e_k)\overline{P_u}$ \\ \hline
$W_i$ & Waiting cost & Arises from the nature of battery charging, which requires a certain amount of time to complete. According to \cite{trans23}, the waiting time varies depending on the type of station and the vehicle's charging capacity, typically ranging from 20 minutes to 50 hours. & $W_i = \overline{w}r(x_{2,i} + x_{3,i})$ \\ \hline
$T_i$ & Travel expenses & Travel cost is defined as the cost incurred by customers traveling from residential areas to stations. According to \cite{luo20}, travel cost is calculated as the product of the per-kilometer access fee, the distance from RP $i$ to station $j$, and $\#$vehicles visiting station $j$ and the penetration rate $r$.& $d_{i,j} P_{km}r\frac{\sum_{k\in\{2,3\}}x_{k,i} e_k}{\overline{e}}$ \\ \hline
\end{tabular}%
}
\label{tab:factor}
\end{table*}

According to \cite{wang20}, the charging time is calculated by \eqref{eq:charging_time}:

\begin{align}
t_\text{Charging} = \frac{e_{100}{\overline{d} } }{100P_{i}} .   
\label{eq:charging_time}
\end{align}

In which ${\overline{d}}$ is average daily driving distance, $e_{100}$ is energy consumption per 100 km and $P_{i}$ is the charging power of the station $i$. Additionally, by incorporating the power factor of $0.85$, this formula provides a more accurate estimate of the actual charging time as \eqref{eq:charging_time085} :

\begin{align}
t_\text{Charging} = \frac{\text{Power Consumption}}{0.85P_{i}}.
\label{eq:charging_time085}
\end{align}

\subsection{Constraints}

The established constraint is regarding the budget for building stations: the total installation cost must not exceed the allocated installation budget $B$:

\begin{equation}
\begin{split}
 c_1: \sum_{i=1}^n \sum_{k\in\{2,3\}} \left( {x_{k,i}}i_{k,i} \right) \leq B,
\end{split}    
\label{eq:c1}
\end{equation}

for each location $i$, $i_{2/3,i}$ denotes the installation cost of a $\text{station}_{\text{Level-2/3}}$, and $x_{2/3,i}$ represents $\#\text{station}_{\text{Level-2/3}}$. Second, the distance between a RP $j$ and a station $j$ must not exceed the maximum driving range \cite{zhou22}:

\begin{equation}
\begin{split}
c_2: d_{i,j} \leq d_{max}.
\end{split}    
\label{eq:c2}
\end{equation}

The demand must be met at each point $i$ to ensure that all charging needs at point $i$ are satisfied. This prevents any shortage of charging capacity relative to actual demand:

\begin{align}
    &c_3: r\sum_{k\in\{2,3\}}x_{k,i}e_{k,i} \geq \bar{e} c_i \label{eq:c3}, \\
    &c_4: r\sum_{k\in[2,3]} (e_{k,i} \sum_{i=1}^{n} x_{k,i}) \geq \bar{e} \sum_{i=1}^{n} c_i. \label{eq:c4}
\end{align}

Constraints $c_3$ and $c_4$ ensure that the charging capacity of the system is sufficient to meet the demand at both individual locations and across the entire network. Constraint \eqref{eq:c3} guarantees that the total charging capacity at a specific location $i$, calculated as the combined energy consumption of Level-2 chargers and Level-3 chargers, is greater than or equal to the energy demand generated by the number of electric vehicles in that area. Constraint  \eqref{eq:c4}  extends this logic to the entire system, requiring the total charging capacity of all Level-2 chargers and Level-3 chargers to be sufficient to satisfy the total energy demand across all locations. Key parameters include $e_{2,i}$ and $e_{3,i}$ (energy consumption per charger), $r$ (charger utilization rate), $x_{2,i}$ and $x_{3,i}$ (number of chargers), $\bar{e}$ (average energy required per vehicle per day), and $c_i$ ($\#$electric-vehicles in a residential area). Together, these constraints ensure a balance between supply and demand, preventing shortages and optimizing the performance of the charging network.

\section{Experiments}
\label{sec:data}

\begin{figure}
    \centering
    \includegraphics[width=0.6\linewidth]{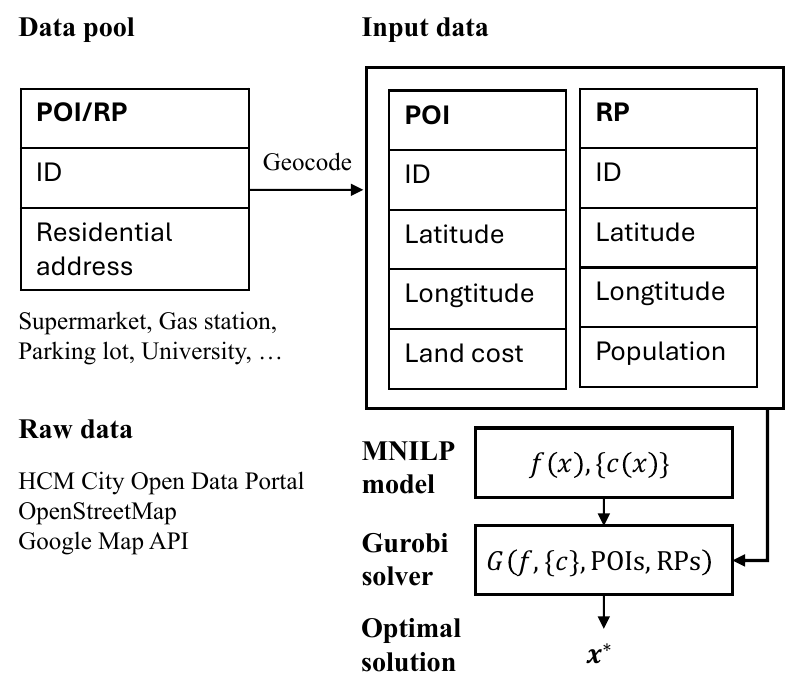}
    \caption{ Experiments schedule with the crawled dataset from Section.~\ref{sec:data}, passed through Gurobi solver.}
    \label{fig:schedule}
\end{figure}

\subsection{Dataset}

Fig.~\ref{fig:schedule} illustrates the data processing and optimization workflow for determining electric vehicle station locations using the MINLP model. Raw data is collected from 3 sources, including OpenStreetMap, Google Map API, and the HCM City Open Data Portal, then processed through geocoding to convert it into coordinates (latitude and longitude). The input data is organized into two tables: Point of Interest (POI) (including ID, coordinates, and land cost) and RP (including ID, coordinates, and population). The MINLP model utilizes this data to define the objective function and constraints, which are then solved using the Gurobi Solver. The process concludes with an optimal solution that determines the locations and quantities of stations to best meet the model's requirements.

In demand modeling, the population is often used as a representative of that region, \cite{batta19} divided the study area into smaller regions, assuming that residents concentrate at the region’s center, in which the area is divided into regions based on estimated EV density and charging demand, with distributed stations ensuring access for all EV users \cite{batta19}. Thus, this study uses population data by ward to represent demand in each area, with each ward centered around its local administrative office, it is also called Residential Point (RP). As of December 31, 2022, the city has 22 district-level administrative units (including 5 rural districts and 1 municipal city) and 249 wards \cite{ctkhcm23}, the POIs and RPs are distributed not equally between districts, as shown in Fig.~\ref{fig:number}. Table \ref{tab:poi} summarizes the 1,509 POIs used as input for the research model, categorized into six main groups: parking lots, supermarkets/shopping malls, apartments/office buildings, universities/colleges, hotels, and gas stations. Among these, apartments/office buildings have the highest number (521 locations, 34.5\%), followed by gas stations (514 locations, 34\%). Places such as hotels (61 locations, 4\%) and parking lots (94 locations, 6.2\%) account for smaller proportions, reflecting an uneven distribution. The diverse dataset, ranging from public spaces to commercial areas and existing infrastructure, highlights that the model has carefully considered various aspects to optimize the placement of stations.

\begin{table}[]
\caption{Statistics of POI Data}
\label{tab:poi}
\centering
\renewcommand{\arraystretch}{1.2} 
\begin{adjustbox}{center, width=0.99\textwidth}
\begin{tabular}{|p{3cm}|p{6cm}|p{1.7cm}|p{2cm}|} \hline
\textbf{Category} & \textbf{Label} & \textbf{Total Number} & \textbf{Source} \\ \hline  
Supermarket, Shopping Mall & market, supermarket, grocery store, hypermarket, mall & 128 & Google Maps \\ \hline  
Gas Station & pvoil, gas station & 514 & \cite{pvmap24} \\ \hline  
Parking Lot & parking & 94 & Google Maps \\ \hline
Apartment, Office Building, Tower & apartment, office building, office space, tower, business tower & 521 & Google Maps, \cite{3292} \\ \hline
University, College & university, college, higher education institution, academic institution & 191 & Google Maps \\ \hline
Hotel & hotel, accommodation, lodge, resort & 61 & Google Maps \\ \hline
\end{tabular}
\end{adjustbox}
\end{table}

\subsection{Metrics}
\indent To evaluate the reliability of the model, three main factors need to be considered: the level of optimization, the solution time, and the feasibility of the solution. The MIP Gap parameter in Gurobi controls the minimum quality of the solution by setting an upper bound on the acceptable gap between the best-known solution and the optimal solution, according to \cite{gurobi20}, is calculated by $\frac{|\text{Obj Bound} - \text{Incumbent Obj Value}|}{ |\text{Incumbent Obj Value}|}\times 100\%$.
 
The smaller the MIP Gap value, the closer the algorithm has found a solution to the optimal value (ranging from $10^{- 6}$ to $10^{- 4}$). The problem is solved when satisfying the above conditions which return feasible solutions in fixed time. Second, execution time is a crucial metric for evaluating efficiency. Finally, feasibility involves verifying that the model meets all its constraints. In other words, the proposed solution must fulfill all specified conditions and limits to reach feasibility. 

\subsection{Results}

\begin{figure}
    \centering
     \caption{Distribution of station locations (a) proposed (\textcolor{blue}{blue} dots) and (b) existing (\textcolor{red}{red} dots) .}
    \includegraphics[width=0.99\linewidth]{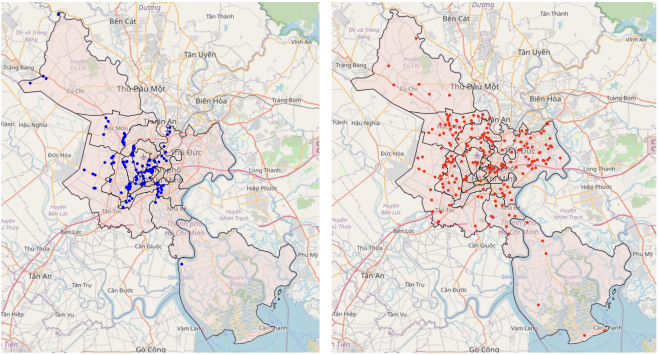}
    \label{fig:red}
\end{figure}

Potential locations were calculated to satisfy consumer demand and minimize costs for both operators and consumers, as shown in Fig.~\ref{fig:red}, existing stations are represented by \textcolor{blue}{blue} dots, while the optimal locations for new stations are shown as \textcolor{red}{red} dots. The both map illustrates the distribution of existing and new EV stations, with a significant concentration in central areas such as \{District 1/3, Binh Thanh District, and Phu Nhuan District\}, where population density and commercial activities are high. This clustering effectively meets the needs of residents and tourists in these bustling areas. However, suburban districts like \{Cu Chi District, Hoc Mon District, Binh Chanh District, and Nha Be District\} have fewer stations, which can create challenges for EV users traveling to these more remote regions.

Notably, the number of new charging points is approximately half of the existing ones. Consequently, in planning, it is crucial to prioritize both the expansion of new stations and the maintenance of existing ones, ensuring continuous surveys, upkeep, and repairs to improve station utilization, especially in areas with less developed infrastructure. This approach will improve the efficiency of stations and ensure equal access to EVs charging infrastructure across all regions.

\begin{figure}
    \centering
     \caption{$\#\text{station}_{\text{Level}-2}$    and $\#\text{station}_{\text{Level}-3}$ by Administrative Unit.}
    \includegraphics[width=0.8\linewidth]{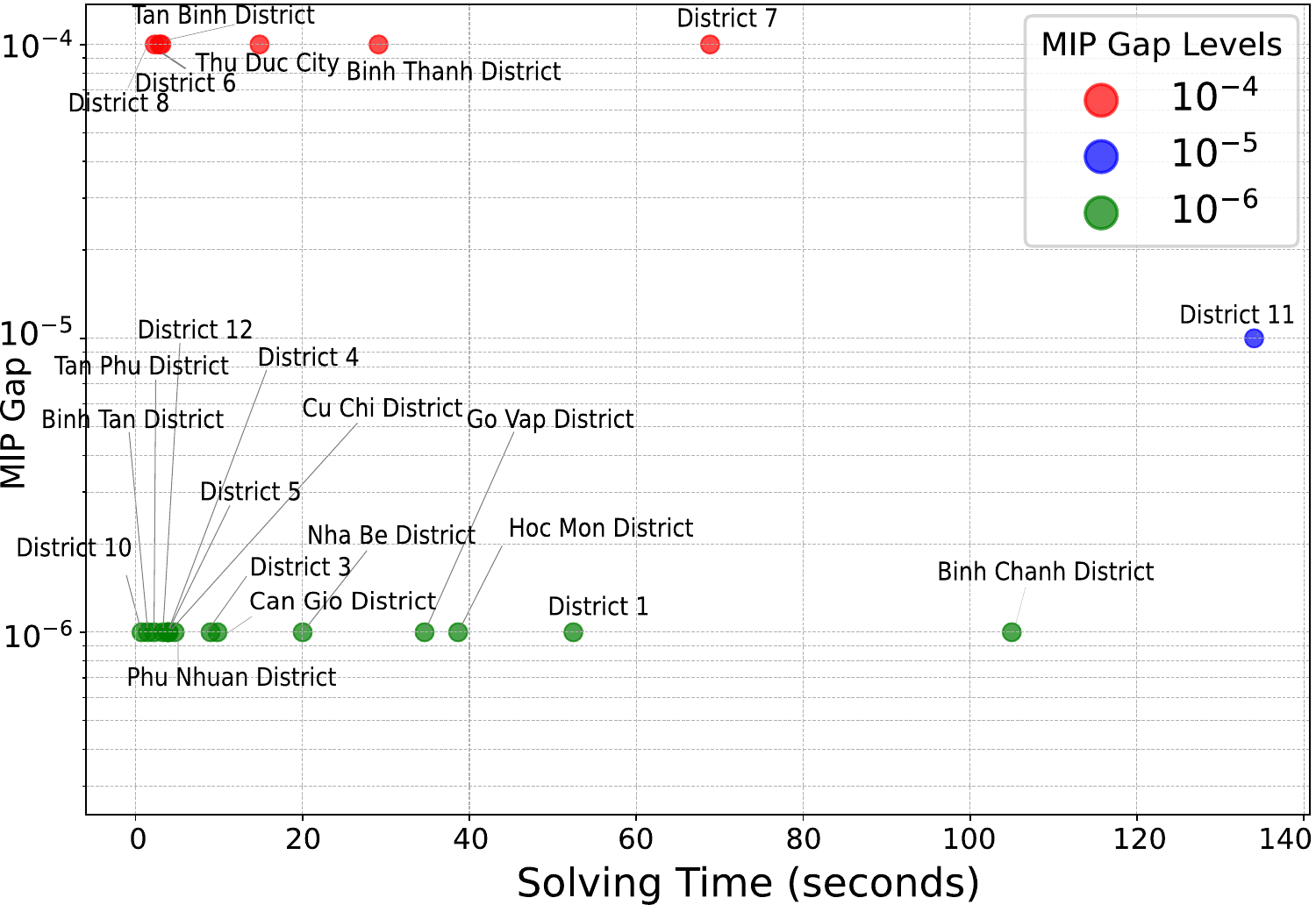}
    \label{fig:solvingtime}
\end{figure}

As presented in Fig.~\ref{fig:solvingtime}, the optimization model applied across all districts achieves a MIP Gap smaller than \(10^{-4}\), demonstrating a high level of convergence for all cases. Based on this, the results are classified into three groups: \(10^{-4}\), \(10^{-5}\), and \(10^{-6}\). Group 1, with a MIP Gap of \(10^{-4}\), includes \{Thu Duc City, Districts 6/7/8, Tan Binh District, Binh Thanh District\}. Group 2, characterized by a MIP Gap of \(10^{-5}\), consists solely of \{District 11\}. Finally, Group 3, featuring the smallest MIP Gap of \(10^{-6}\), represents the areas with the highest precision and convergence, encompassing 15 districts: \{Districts 1/3/4/5/10/12, Go Vap District, Tan Phu District, Phu Nhuan District, Binh Tan District, Cu Chi District, Hoc Mon District, Binh Chanh District, Nha Be District, and Can Gio District\}. The predominance of the \(10^{-6}\) group highlights the model's superior accuracy and scalability in these regions, underscoring its robust performance.

Furthermore, the results from Fig.~\ref{fig:solvingtime} show the solving time (in seconds (s)) of the optimization model categorized by administrative units, revealing significant variation in computational efficiency across different areas. The first group, with solving times ranging from $1.16$ to $6.95$ (s), includes \{District 4/6/10, Tan Binh District, Binh Thanh District, Hoc Mon District, Can Gio District\}, highlighting these areas as the most computationally efficient. The second group, with solving times between $9.5$ and $15.81$ (s), consists of \{Thu Duc City, District 3/8, Binh Chanh District\}, showing moderate solving times. The third group, with solving times spanning $20.18$ to $44.39$ (s), includes \{District 1/7, Phu Nhuan District, Binh Tan District, Nha Be District\}, indicating a higher computational demand. Finally, the fourth group, with solving times ranging from $48.42$ to $99.47$ (s), represents the least efficient areas, comprising \{District 5/11/12, Go Vap District, Tan Phu District, Cu Chi District\}. This progression suggests that solving time increases with the complexity or size of the problem in these regions, potentially reflecting differences in infrastructure, data characteristics, or model scalability.

\begin{figure}
    \centering
     \caption{Distribution of $\text{station}_{\text{Level-2}}$ and $\text{station}_{\text{Level-3}}$ by Administrative Unit.}
    \includegraphics[width=0.8\linewidth]{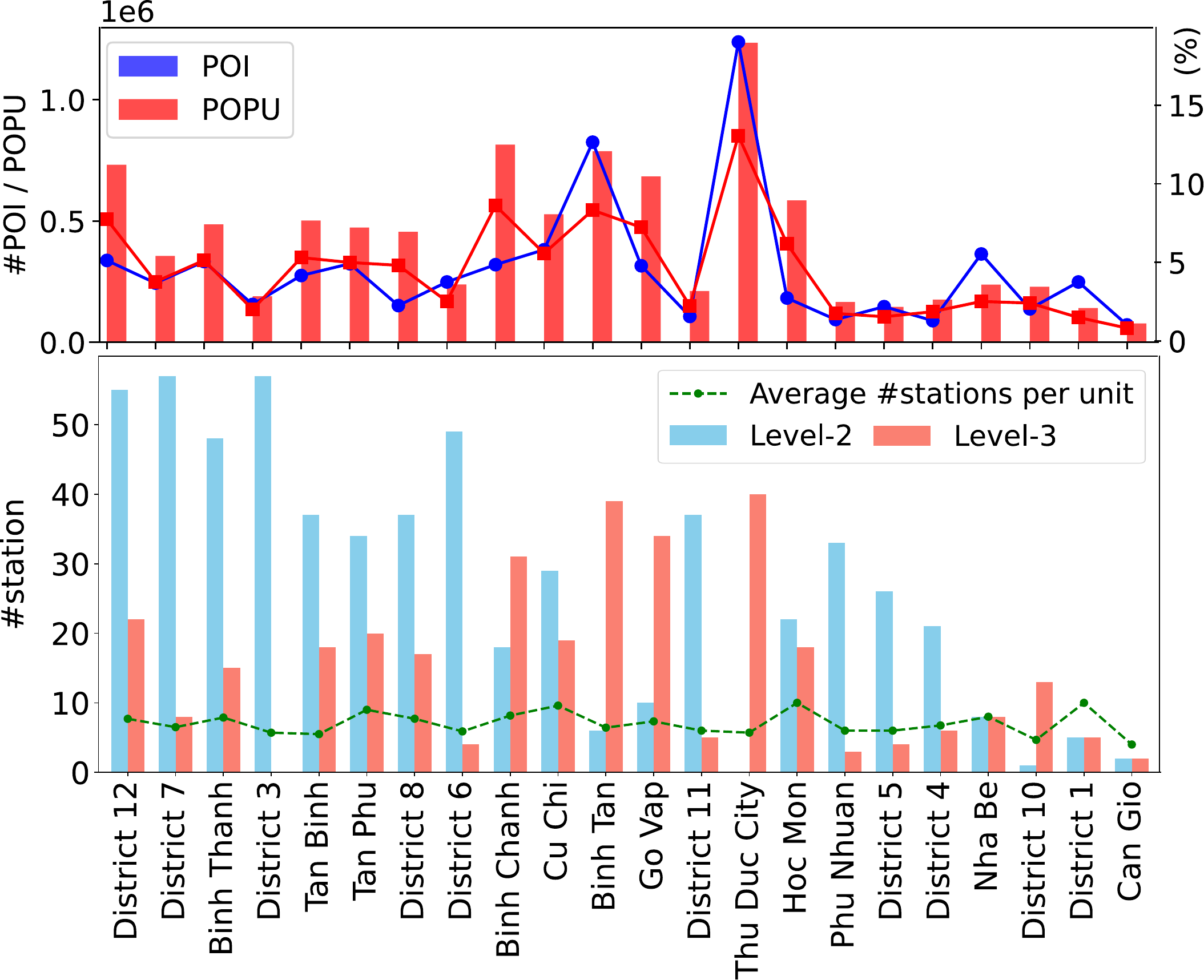}

    \label{fig:number}
\end{figure}

\begin{figure*}
    \centering
     \caption{(a) Population density at district level As of December 2023, HCM City, covering just 0.63\% of Vietnam's total area (2,095 $\text{km}^2$), is home to about 9 million residents, making up 10\% of the nation's population. (b) The selected RPs are marked by (\textcolor{green}{green} stars), with each star corresponding to a chosen local administrative office (c) EV stations (\textcolor{blue}{blue} dots) optimal solution distribution.}
    \includegraphics[width=0.99\linewidth]{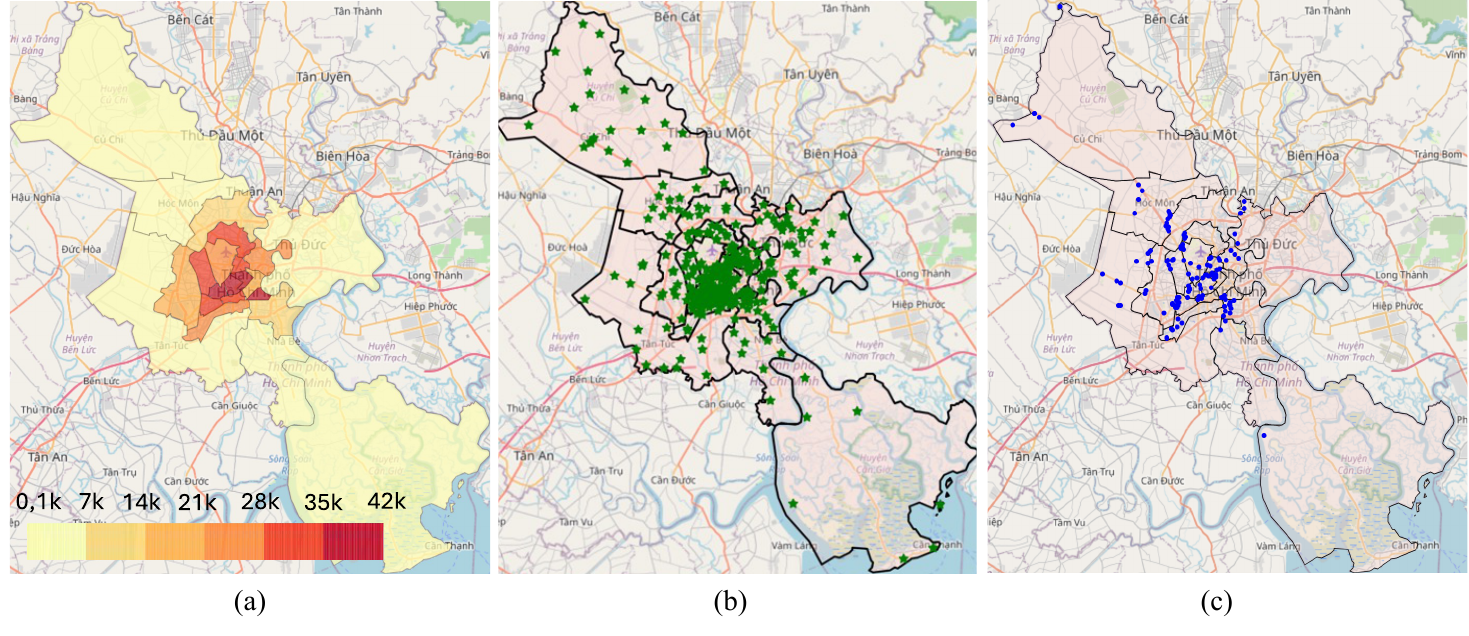}
    \label{fig:density}
\end{figure*}

Fig.~\ref{fig:density} illustrates the strong connection between population distribution, RP, and the optimal placement of EV stations in HCM City. Fig.~\ref{fig:density}(a) shows population density, with central districts such as \{District 1/3, Binh Thanh District\} having the highest densities, exceeding $42,000$ people/$\text{km}^2$, while suburban areas like Can Gio have much lower densities. Based on this data, Fig.~\ref{fig:density}(b) identifies RPs, which are predominantly distributed in densely populated areas such as \{Tan Binh, Go Vap, District 7\} while ensuring coverage across the entire city. Fig.~\ref{fig:density}(c) presents the optimal solution for EV station placement, with stations concentrated in high-demand areas like \{District 1/5 Binh Thanh District\}, while also extending to suburban areas such as Hoc Mon and Nha Be to meet long-distance travel needs. The optimization model effectively leverages data on population density and RP distribution to design a charging network that addresses high demand in central districts while ensuring accessibility for suburban residents. This approach balances supply and demand, enhancing the feasibility of EV charging infrastructure in both urban and suburban areas of HCM City.

\section{Insight}

Fig.~\ref{fig:number} (Top) highlights the distribution of POIs and average population across administrative units in HCM City, reflecting the relationship between infrastructure and population demand. Thu Duc City has the \textbf{highest} number of POIs (278 points, 19.02\%) and the \textbf{largest} population (13.06\%), indicating its strong potential for EV charging infrastructure development. \{District 12, Binh Thanh District, Binh Tan District\} also show high POI and population ratios, aligning with real-world demand. In contrast, central districts such as \{District 1/3\}, despite having lower populations, still have a relatively high number of POIs due to their roles as economic and commercial hubs. Suburban areas like Can Gio and Nha Be have lower POI counts and populations, suggesting the need to expand charging infrastructure to improve accessibility for residents. Overall, the current POI distribution aligns reasonably well with population density, but further optimization is needed to better meet demand in both central and suburban areas of HCM City.

Fig.~\ref{fig:number} (Bottom) show the newly installed stations (including Level-2 stations and Level-3 stations) are concentrated in central urban districts and suburban areas lacking infrastructure, with over 50 stations needed across 8 districts. Leading the demand is District 12 (77 stations), followed by District 7 (65 stations) and Binh Thanh District (63 stations). These areas show high demand for EV charging, driven by rapid economic growth and key development zones in District 12 and the expanding economic area in District 7. Conversely, \{District 1, Can Gio\} have the lowest numbers, with only 4 and 10 stations respectively. Can Gio District's low density and limited urbanization explain the low demand, while District 1's developed infrastructure meets existing needs, reducing the need for new installations. 

Overall, Level-2 stations outnumber Level-3 stations, except in {Binh Chanh District, Binh Tan District, Go Vap District, Thu Duc City, District 10}. Notably, Thu Duc City, despite having the highest $\#$POI, has no Level-2 stations. Regarding POI and RP, while the Regional Priority is evenly distributed, \{Thu Duc City, Binh Tan District\} stand out with numerous $\#$POI, highlighting significant opportunities and potential in these areas. The results indicate that the factors influencing cost optimization for operators and station owners are ranked in descending order of impact as follows: Charging cost, Operational cost, Travel cost, Waiting cost, Installation cost, Land rental cost and Maintenance cost. Among these, the charging cost has the \textbf{most} significant influence, accounting for an average of \textbf{35.47\%} of the total charging costs. This is followed by the operational cost, which contributes \textbf{28.83\%}, and the travel cost which accounts for \textbf{27.99\%}. In contrast, the factor with the least average impact is the maintenance cost representing just \textbf{0.27\%}. Charging costs are the largest expense in the EV model, emphasizing the need for reducing the fees that EV users have to pay. This can be adopted by short-term promotions during peak hours and long-term strategies, with the Vietnam Electricity Group (EVN), the state-owned monopoly responsible for electricity distribution and pricing, playing a central role. EVN should consider implementing preferential electricity pricing policies for electric vehicle charging stations. Next is operational cost, investments in technologies like Vehicle-to-Grid (V2G) systems \cite{zhang22_1}, battery swapping \cite{agha22}, and renewable energy integration \cite{tan23} is needed to diversity energy source. Besides, travel costs are significant, particularly in remote areas such as \{Can Gio District, Cu Chi District, Binh Chanh District, Thu Duc City\}, where travel expenses may exceed 50\% of the total cost, requiring real-time station updates (e.g., availability, maintenance schedule) to enhance convenience. Additionally, to address waiting costs, operators should use mobile apps for real-time availability and improve facilities with amenities namely cafes, convenience stores, and restrooms to enhance the user experience. Meanwhile, installation costs demand government collaboration, safety compliance, and urban planning support, while public land use and multi-level station designs can reduce land rental expenses \cite{aayog21}. Subsequently, regular preventive maintenance with predictive technology ensures station reliability, safety, and longevity while minimizing repair costs.

\section{Conclusion}
In an era of rapid technological advancement, the demand for EVs and thus, charging infrastructure requests is growing expeditiously, requiring businesses to continuously expand charging networks to meet user needs, especially amid the shortage of public stations in Vietnam. Optimizing EV charging systems has thus become a state-of-the-art and essential approach to support this increasing shift towards electric mobility in the country. This study was conducted to enable informed decisions about optimal locations for stations to be placed as well as identifying the factors influencing the deployment and operation of station systems. It synthesized previous research to build a rigorous model incorporating seven cost factors: installation, land rental, maintenance, operational, charging, waiting, and travel. Results from the Gurobi solver revealed that charging, operational, and travel costs make up a significant portion of the total expenses. Finally, the author discussed feasible solutions to enhance the operational efficiency of EV stations for different user groups and highlighted the study’s limitations, offering insights to guide future research in refining the optimization model for EV charging networks.

Currently, Vietnam is in the process of developing regulatory standards for electric vehicle charging stations. In the future, station construction will need to comply with these standards, meaning that not all potential POIs will be eligible for installation. Therefore, future studies should incorporate these evolving regulations when selecting station locations to ensure models remain relevant and applicable. Additionally, expanding the scope of research to include multiple provinces across Vietnam would provide a more comprehensive and representative analysis. Beyond geographical considerations, future research should also integrate dynamic EV usage patterns and user preferences, as this study primarily relies on static data. Moreover, sensitivity analysis using different algorithms would help assess model robustness and computational complexity, enhancing its applicability in diverse scenarios. Lastly, if time and resources permit, incorporating expert insights from relevant fields could further refine the study, improving both the depth and reliability of the findings.

\section*{Code and Data availability}

The data that support the findings of this study are openly available at https://github.com/tulliavu/EVCSVN.

\section*{Acknowledgment}

The authors thank the Gurobi team for providing an Academic license for the Gurobi optimizer.



\bibliographystyle{IEEEtran}
\bibliography{ref.bib}

\end{document}